 \documentclass[final,3p,times,twocolumn]{elsarticle}
\usepackage{enumerate}
\usepackage{flushend}
\usepackage{hyperref}

\graphicspath{{../../figs/}}

\begin{document}

\begin{frontmatter}

\title{The Under-Water Dark-Room Experimental Facility at the University of Winnipeg}
\author{  Ajmi Ali}\ead{a.ajmi-ra@uwinnipeg.ca} 

\author{\normalfont Blair Jamieson, Lyndsay Green, Tapendra BC, Rituparna Banerjee, Mahnoor Mansoor, Andrea Mayorga, Anna Harms, Fabio Castellanos Lenes, Brijesh Sharma, Flora Easter, David Ostapchuk, Shomi Ahmed, Kyle~Macdonald, Craig Wood, Marshall Kirton, Gonzalo Paz}
\affiliation{organization={Department of Physics, University of Winnipeg},
            addressline={515 Portage Avenue}, 
            city={Winnipeg},
            postcode={R3B2E9},   
            state={Manitoba},
            country={Canada}}

%
%
 

\begin{abstract}

 A completely new  under-water dark-room test facility  (UWDTF) has been built at the University of Winnipeg during 2021-2023, for the testing of the equipments, optical components and detectors before they might be used in different underwater experiments, like the Hyper-Kamiokande (Hyper-K), and others. The Facility is designed for Research and Development activities primarily related to the different calibration systems, which are/will be used in the Water Cherenkov Test Experiment (WCTE) at CERN, the Intermediate Water Cherenkov Detector (IWCD) at Tokai, Japan and the Hyper-Kamiokande Far Detector at Kamioka, Japan.
 The facility houses a large tank of water (1000 gallons) in an optically isolated room, and is equipped with a gantry that provides for the 3D motion of a maximum of 50 lbs of load inside the tank. A customized pan-tilt system has also been devised to accommodate further degrees of freedom of motion to the payload in the polar and azimuthal direction.
 The facility is primarily used for testing of the under-water camera housings designed for the Hyper-K experiment, besides many other research and development activities. The preliminary results of the camera calibration done in this multi-purpose underwater-darkroom facility are presented here, starting with the description of the vital features of this facility.  
\end{abstract}



\end{frontmatter}

\section{Introduction}

 HyperKamiokande (Hyper-K) \cite{hk1,ali} is a globally collaborated neutrino long-baseline experiment being built in Japan currently, to determine the CP volation in the lepton sector \cite{Fogli:2012ua, Gonzalez-Garcia:2014bfa, branco, branco2, dick}. It comprises of a huge water target of 260~kton, under a mountain rock covering, and will receive neutrinos from the accelerator source 295 km away at JParc, Tokai \cite{Hyper-KamiokandeProto-:2015xww, Igarashi:2021npv}. It will also detect neutrinos from the atmospheric origin \cite{hk1}. The Hyper-K will determine the values of the neutrino oscillation parameters to utmost precision. The Intermediate Water Cherenkov detector (IWCD) \cite{Scott:2016kdg} is being built closer to the neutrino beam source, to aid in this process. A near detector with the same mode of detection as the water Cherenkov detector at Kamioka, will measure neutrino interactions before oscillation, and contribute to reducing the systematic uncertainties in the measurements at the far detector. A mock-up detector, similar to the IWCD, but smaller in dimensions, called the WCTE (Water Cherenkov Test experiment) \cite{Rodriguez:2024qeq} is hence arranged at the CERN, to validate the working of the various detector components of the future detector.
 
 However, measurements in the smaller water cherenkov detectors themselves also involve uncertainties, which become significant given the smaller size. Therefore various additional calibration methods are necessary in the WCTE and IWCD. One of them is the photogrammetry, which is developed by the members at the University of Winnipeg. Details on this work will soon be presented in future publications. This report mainly documents the details of  building of the facility, which plays a crucial role in the research and development activities, tests and experiments performed for the photogrammetry and other methods required for the Hyper-K project.
 


 A dark-pool facility has been built at the University of Winnipeg with the primary aim of testing and calibration of the components to be used in the WCTE, the IWCD and the Hyper-K Far Detector. The facility houses a large tank of water (1000 gallons) in an optically isolated room to provide for testing of the cameras, the working of the laser system to be used underwater, along with all other electrical components. Several points of consideration had to be incorporated while building such a well-functioning facility within the confined spaces of a university building, and hence are described in the following sections.
 


 
 \section{The Water tank}
 The primary objective of this UWDTF is to study the equipments or perform tests underwater, in an optically isolated environment. The source and the detector system are intended to be as distant from each other, as possible. Therefore, a large accommodating water tank is the central component of this facility, and needs to be carefully installed to stand the needs and wears of time.
 
 Whether customised or commercially available, open-tanks of large dimensions generally require metallic cross braces to support the water pressure due to increased heights above the ground level. So, a tank of optimum dimensions had to be chosen, to suit the purspose of the facility.
 The large dimensions of the tank with complete open access at the top is necessary to allow for unhindered lowering, lifting and moving of the objects into and out of the tank, and also accommodate externally driven motion of the underwater objects for the intended research activities. 
 
\begin{figure}[b!t]
 \centering
\includegraphics[width=1.\columnwidth]{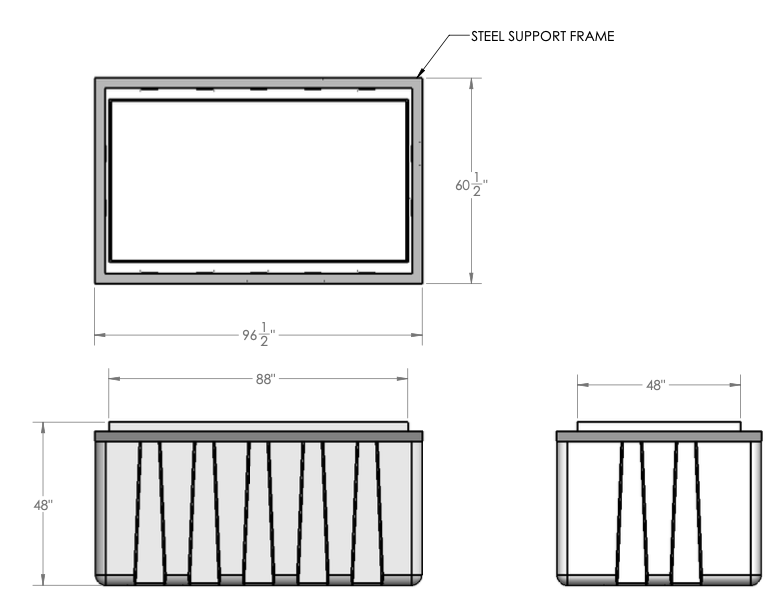}
    \caption{Structural features of the water tank used in the facility.}
\label{fig1}
\end{figure}

 The other important point of consideration in this regard, is the material of the tank. Glass is a well established material used in most aquatic systems worldwide, but given the requirement of potential usage of heavy machinery and metal instruments inside the tank, the material of the tank must be of non-glass material. Metallic tanks would add to the weight of the water load and the different machinery or equipments to be contained inside, and would also add to difficulties in moving and handling of the tank, if and when needed.\footnote{The facility is built in the basement of the University building, after reinforcing the floor with the live load capacity of above 100psf in order to safely support the weight of about 4~tonnes of water across a $8'\times5'$ base-dimensions.} Fortuantely enough, a working option was available in Heavy Density Poly-Ethylene or HDPE. This tank suited closest to the requirements of the facility, and measures $8'\times5'\times4'$ in dimensions, the combination of maximum base cross section and height that could be availed under the given conditions. It has a bolted steel rim, the only metallic structure perimetering the top open surface of the tank. The walls of the tank are carved with the standard trapezoidal shapes to resist the bulging of the walls which support the water pressure at increased heights, as shown in fig.\ref{fig1}.

\section{The Gantry System}

The second most important element of this UWDTF is the  3D-motion gantry with pan-tilt system which would provide for different tests and experiments, like studying the required/specific characteristics of the mPMTs\footnote{mPMT stands for multi-PMT or Photo Multiplier Tubes, a module comprising of 19 PMTs together in an enclosed metal chamber with optically transparent window, the detector unit designed for use at the IWCD.}, testing the performance of the photogrammetry apparatuses, with multiple mPMT modules underwater simultaneously. 

It is necessary to ensure the freedom of motion to the source or the detector harnessed to the gantry arm, in all directions, X, Y and Z axes, as well as rotate in the polar and the azimuthal directions.
 
\begin{figure}[b!t]
 \centering
\includegraphics[width=0.8\columnwidth]{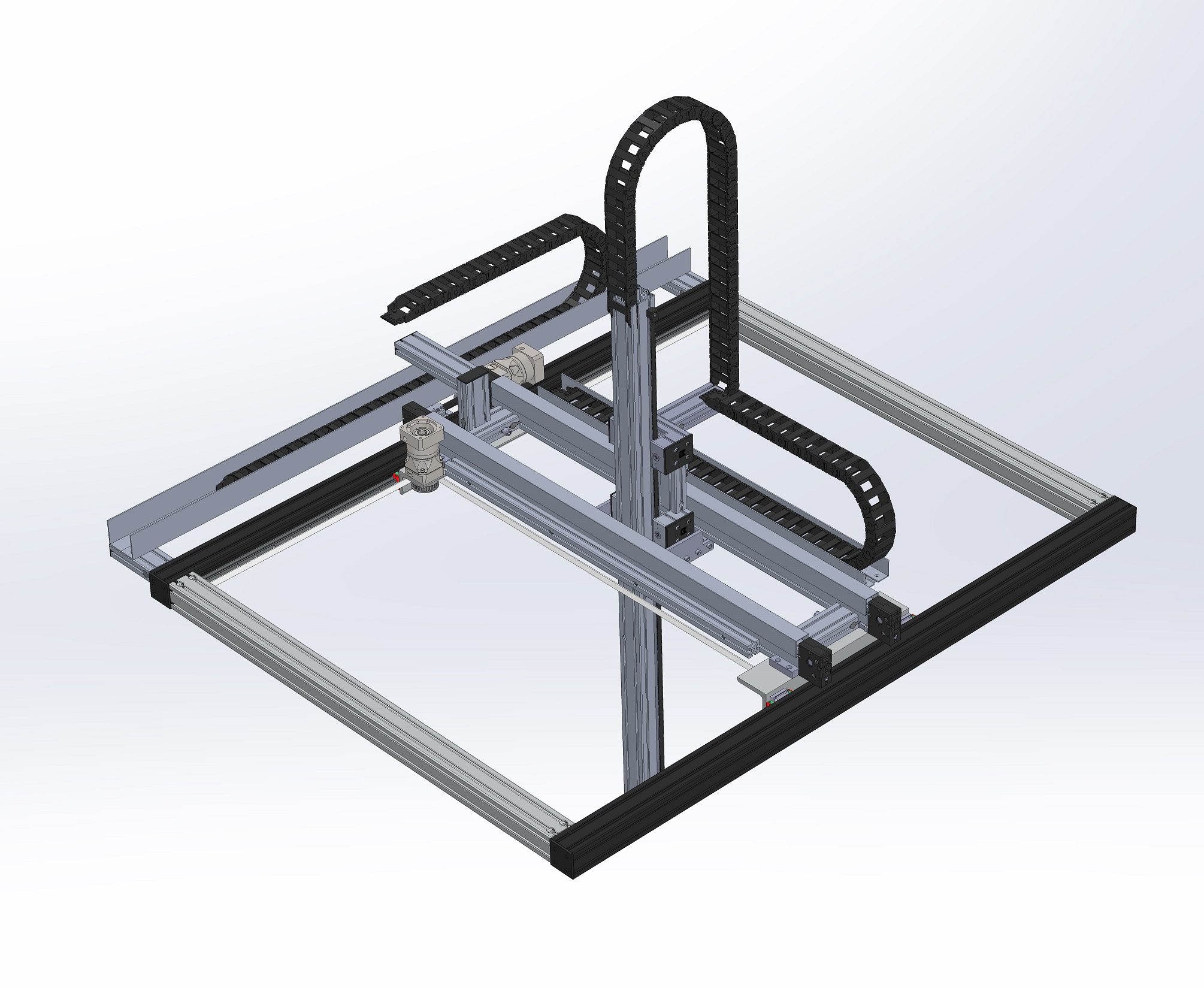}
    \caption{The Cartesian Gantry, with motion along the three axes, X, Y and Z.}
\label{fig2}
\end{figure}

 A cartesian gantry system was designed by the Macron Dynamics shown in fig.\ref{fig2}, customised to the purpose of the facility. The gantry sits on a six-legged structure made of aluminium extrusions fencing the water tank. Each extrusion measures 80~mm $\times$ 80~mm in cross section, a capacity to easily support the gantry along with an additional payload of at least 10~kg. 

 The motion in X-Y-Z directions in the gantry is achieved by three stepper motors. 
 The Z-arm of the Gantry extends down into the water tank, and customised for the room-height, allowing a free vertical motion upto 65~cm. The maximum bending radius of the belt had to be modified to as minimum as 10~cm, 
 to accommodate for maximum vertical motion until the total of 8.5' height of the ceiling in the facility. The supporting structure made of aluminium extrusions just matches the   height of the water tank so as to allow maximum vertical motion.

\begin{figure}[b!t]
 \centering
\includegraphics[width=1.\columnwidth]{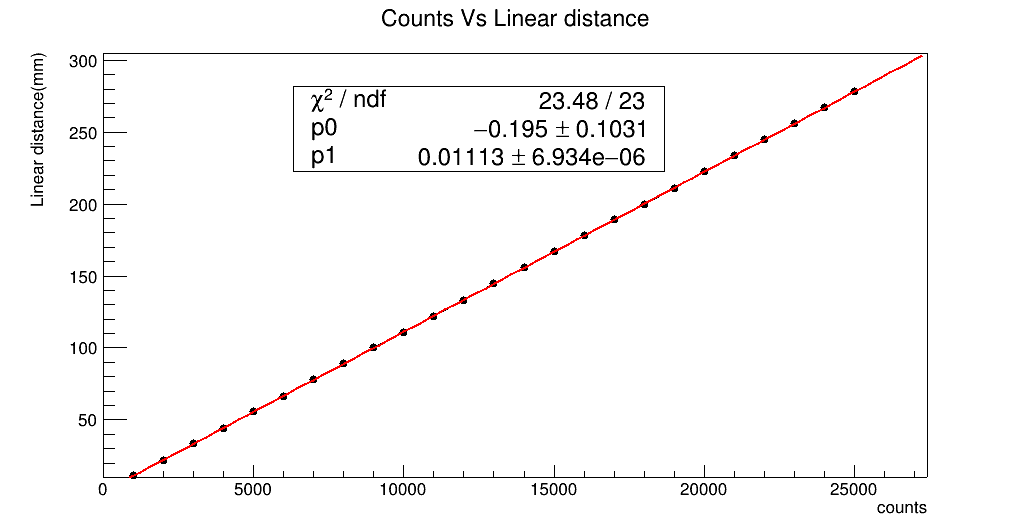}
    \caption{Calibration along the X-axis motion of the gantry.}
\label{fig3}
\end{figure}

\begin{figure}[b!t]
 \centering
\includegraphics[width=1.\columnwidth]{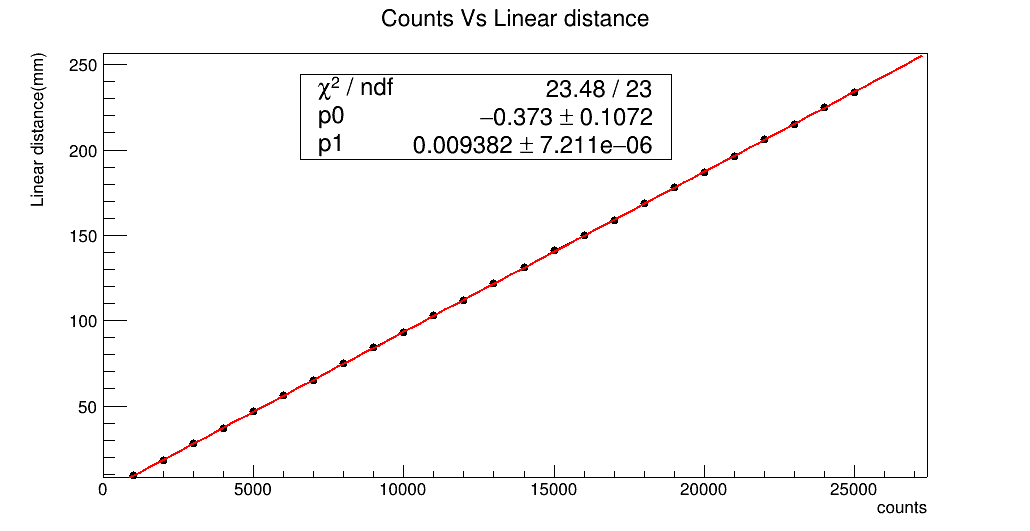}
    \caption{Calibration along the Y-axis motion of the gantry.}
\label{fig4}
\end{figure}

\begin{figure}[b!t]
 \centering
\includegraphics[width=1.\columnwidth]{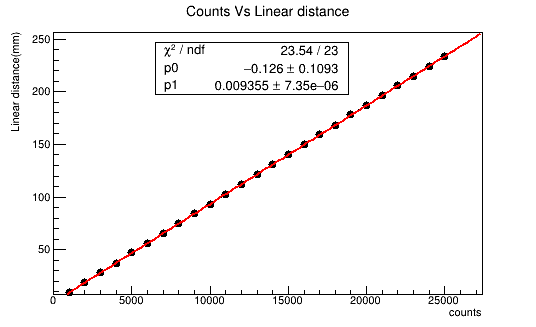}
    \caption{Calibration along the Z-axis motion of the gantry.}
\label{fig5}
\end{figure}

 The XYZ-Gantry system can utilize an entire cubic work envelope of 96\% of their space and size. The three stepper motors are controlled by the Galil motor controller DMC-4153, that can control motions in all the 5 axes independently, be it jogging or point-to-point positioning and many more applications. The distance covered by the gantry motion as dictated by the counts were calibrated along the X, Y and the Z-axes, and are shown in the fig.\ref{fig3}, \ref{fig4}, \ref{fig5}.

\begin{figure}[b!t]
 \centering
 \includegraphics[width=0.38\columnwidth]{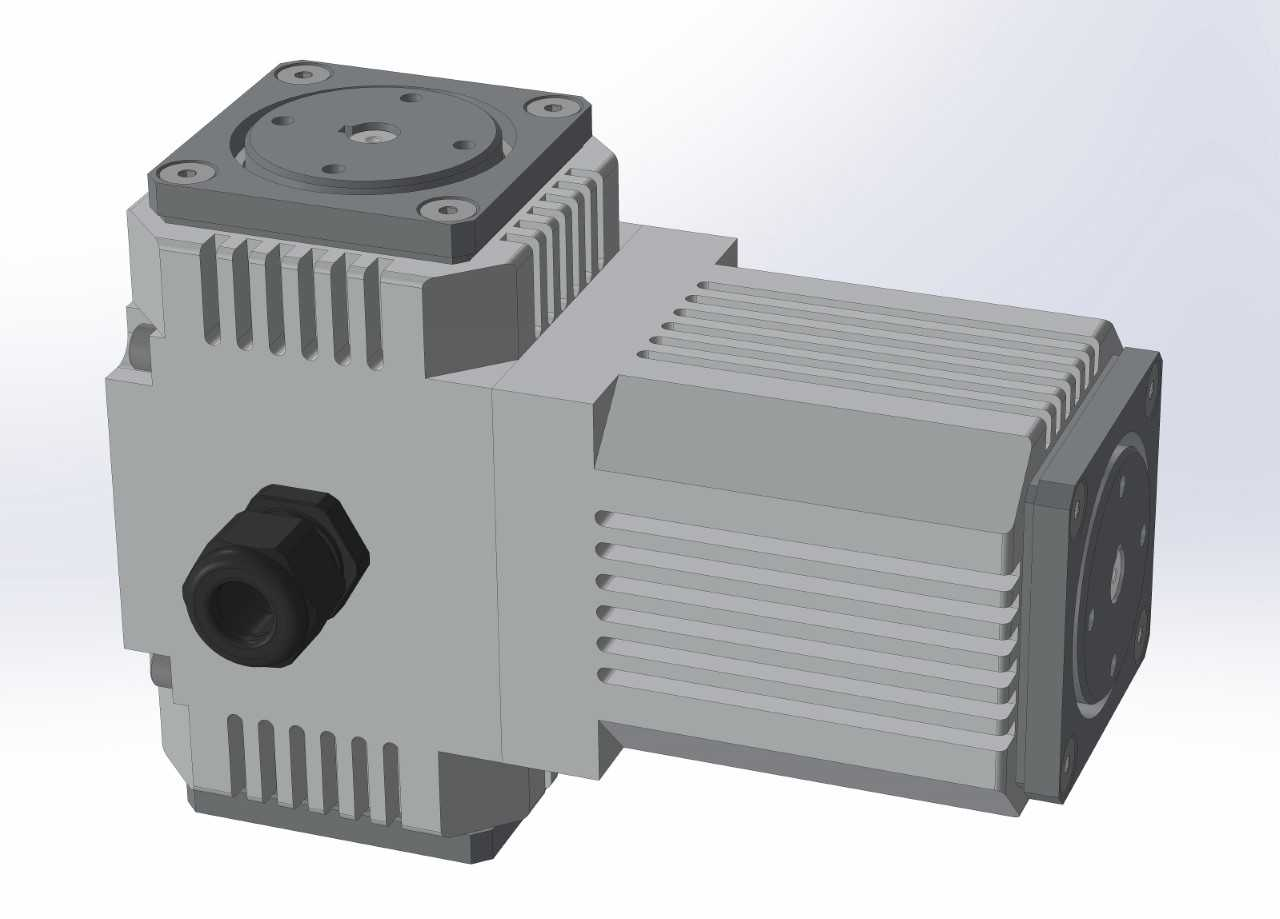}
 \includegraphics[width=0.58\columnwidth]{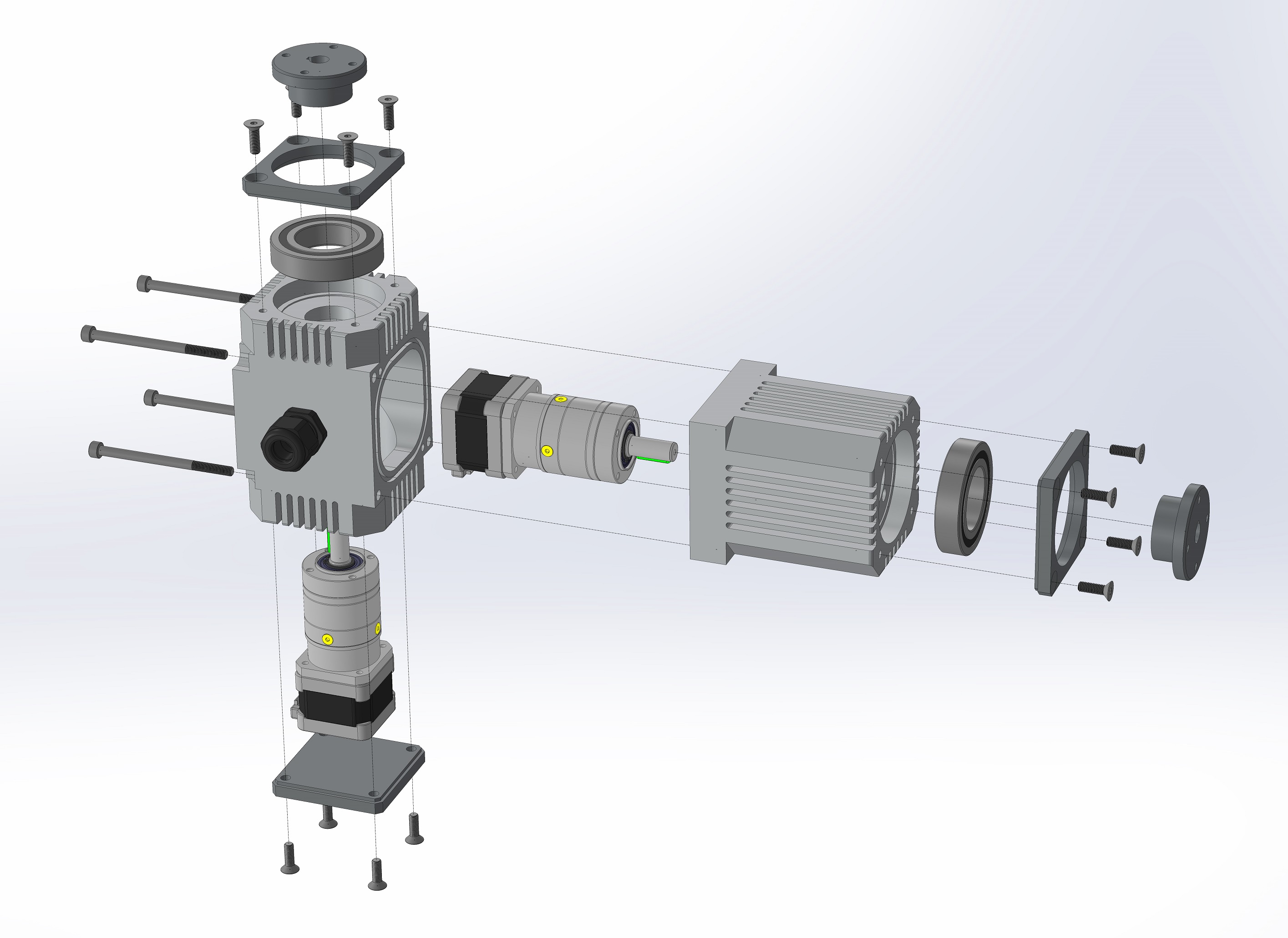}
    \caption{The pantilt system casing on the left, and the blown-up schematics showing the two motors on the right.}
\label{fig6}
\end{figure}

 The payload, be it the source or detector component of the experiment concerned, is attached to the end of the Z-axis, enabling its freedom of movement in all three directions. A pan-tilt system was designed and installed at the end of the Z-axis, with the technical expertise of the Macron Dynamics. The system comprises of two stepper motors, the first attached to the Z-arm, while the second motor sits on the axle of the first, enabling rotation of the payload along the polar and the azimuthal direction. The twin motor-systems are packed inside a water-proof concise casing prepared by the Macron Dynamics, as can be seen in the shematic diagram in fig.\ref{fig6}.
 
 
\begin{figure}[b!t]
 \centering
\includegraphics[width=1.\columnwidth]{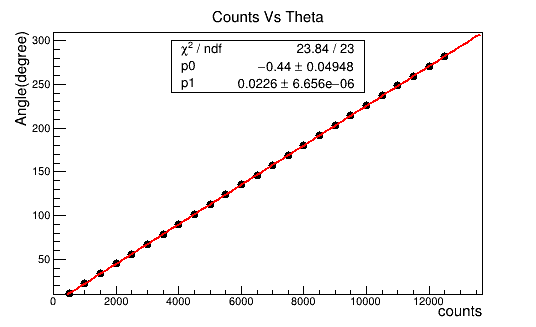}
    \caption{Calibration along the theta-axis or motion in polar direction by the pan-tilt.}
\label{fig7}
\end{figure}

\begin{figure}[b!t]
 \centering
\includegraphics[width=1.\columnwidth]{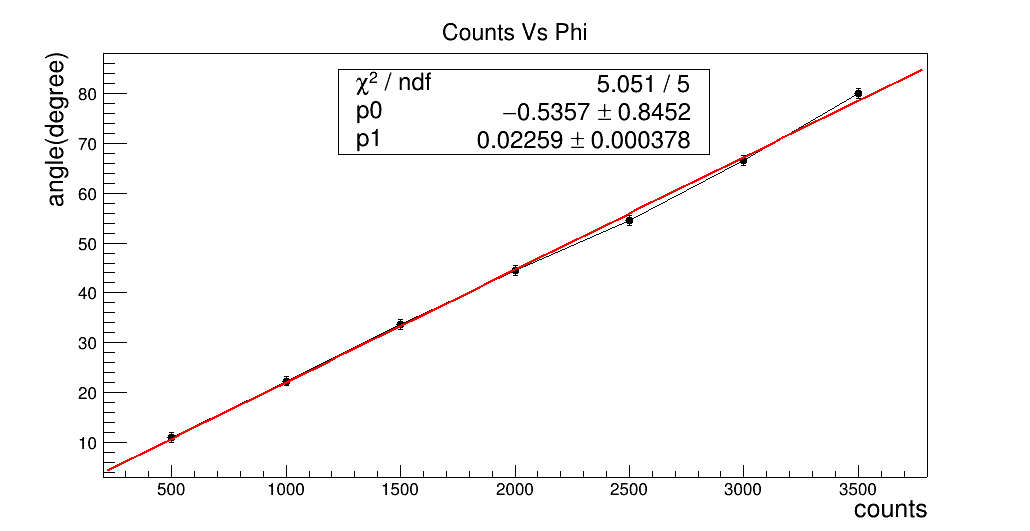}
    \caption{Calibration along the phi-axis or motion in azimuthal direction by the pan-tilt.}
\label{fig8}
\end{figure}

The angles in the polar and the azimuthal direction covered by the pan-tilt motion as dictated by the counts were calibrated and are shown in the fig.\ref{fig7}, \ref{fig8}.

 \section{Operation of the facility}
 
 The dark-room facility with the prime components as described in the earlier sections, is built in the basement of the University of Winnipeg, after reinforcing the floor with the live load capacity of above 100psf in order to safely support the weight of about 4~tonnes of water across a 8'$\times$5' dimension. 
 
\begin{figure}[b!t]
 \centering
\includegraphics[width=1.\columnwidth]{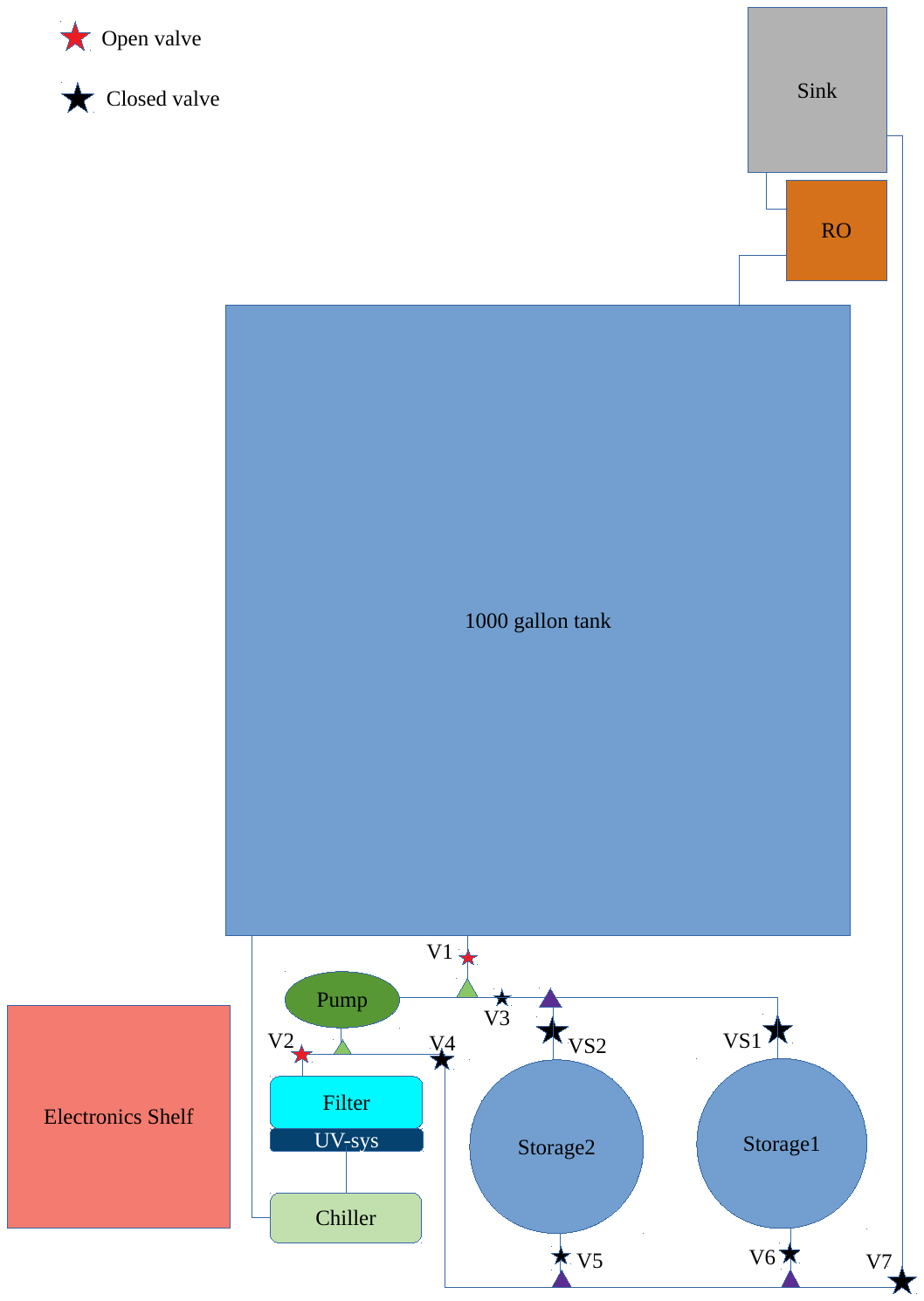}
    \caption{Water Circulation Loop, designed to avoid any microbial contamination in the water system.}
\label{fig9}
\end{figure}

  The facility is mainly meant to house the water tank measuring $\sim$ 5' $\times$ 8', 4' height, and its necessary accompaniments. Two storage tanks of 500 gallons each are also accommodated in the room, to store the water from the open tank, when necessary. The water is filled into the tank/s, through a RO-filter, and is then continuously purified through a regular filter, followed by a UV-filter to avoid microbial contamination in the water-circulation loop, as shown in the fig.\ref{fig9}. A chiller is also deployed in the water-system loop to maintain a steady temperature of around 10$\deg$C of the tank water. This not only ensures a steady temperature during the experiment performed, but also diminishes the possibility of bacterial growth inside the tank and the components.
 
\begin{figure}[b!t]
 \centering
\includegraphics[width=1.\columnwidth]{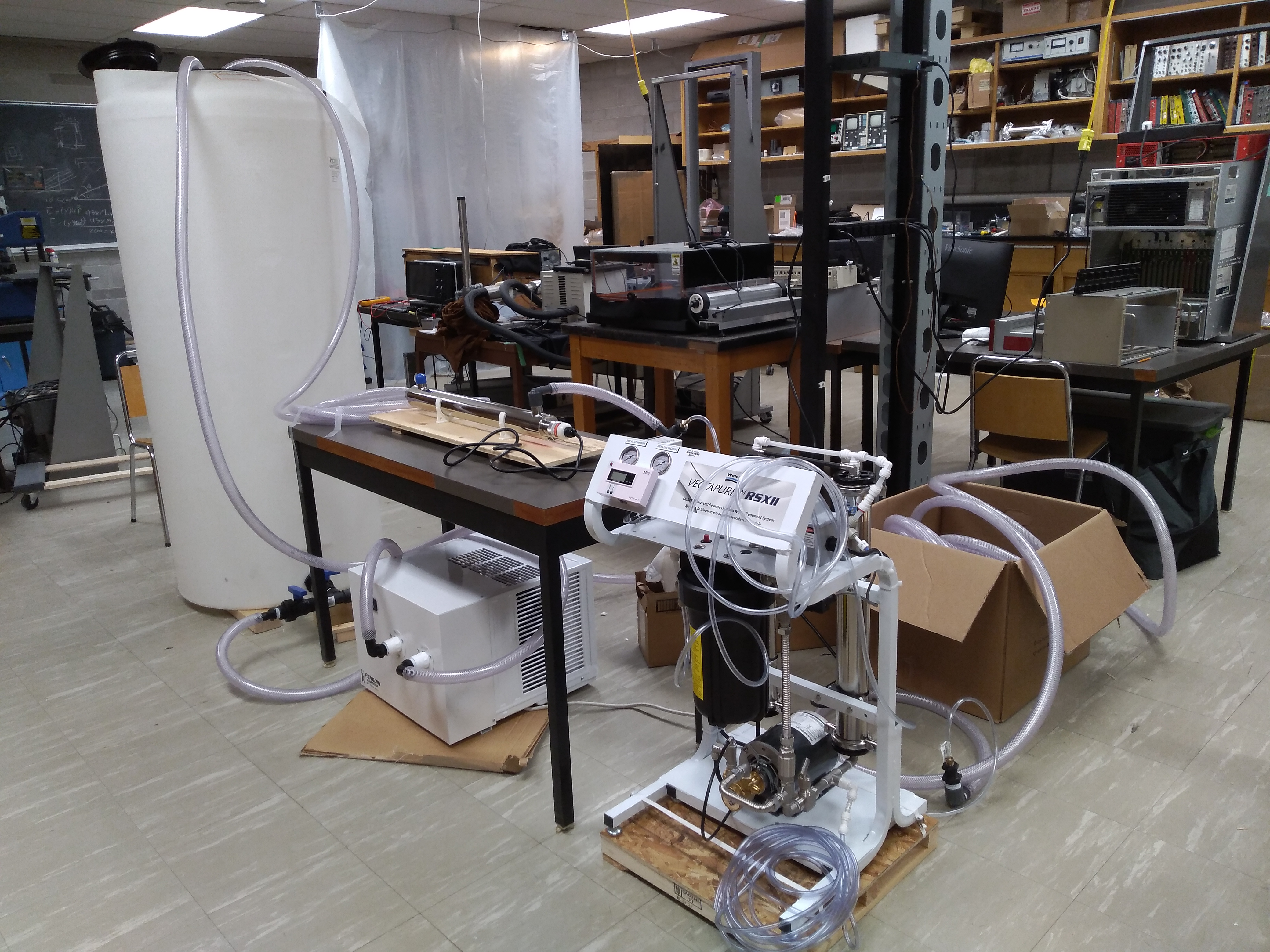}
    \caption{Set-up used for checking the water-circulation system in the laboratory, before installation at the Facility.}
\label{fig10}
\end{figure}

A primary inspection of the water-circulation system was done with a storage tank of 265 gallons capacity in the laboratory, as shown in fig.\ref{fig10}, prior to the installation of the final system with improved pumps at the facility. The RO-filter has the drawback of surplus draining of water, and hence will be replaced with a better alternative in the future.

\begin{figure}[b!t]
 \centering
   \includegraphics[trim=0 2cm 0 6cm, clip=true, width=0.48\columnwidth]{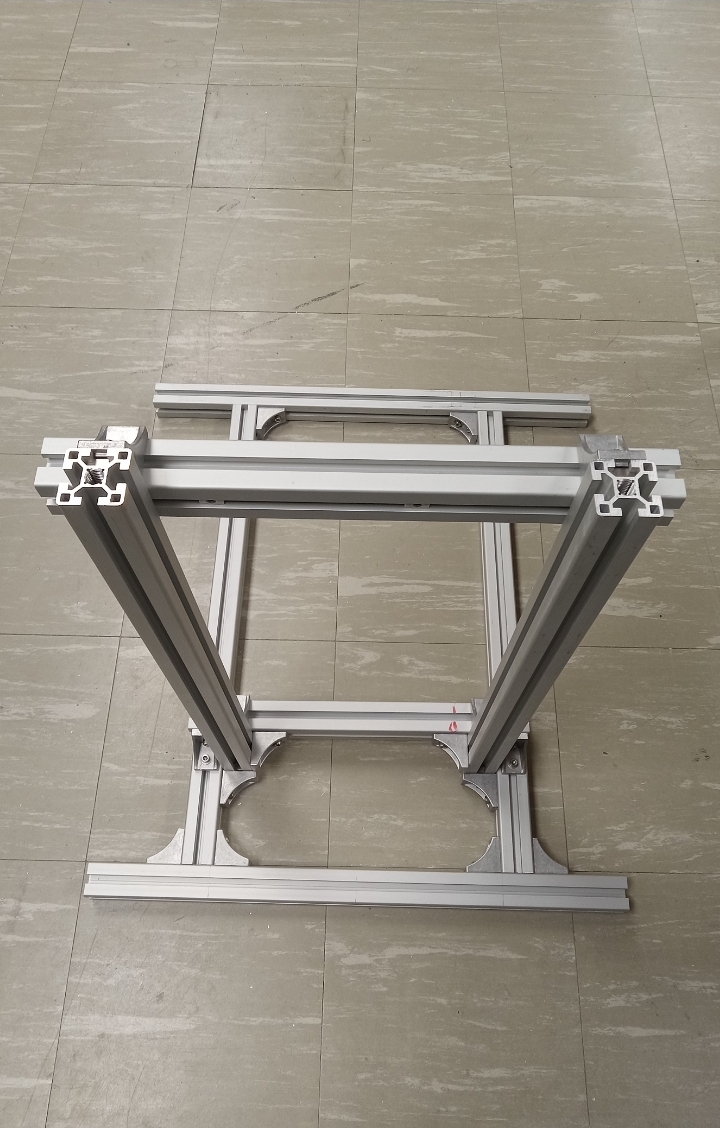}
   \includegraphics[width=0.5\columnwidth]{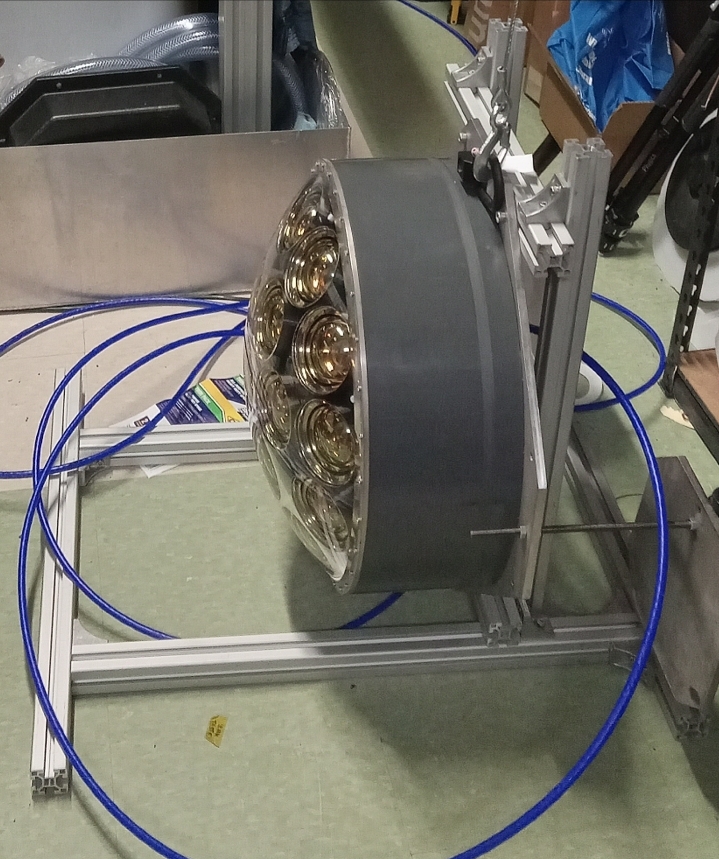}
    \caption{The Detector mounting structure, with two potential mounting positions, the vertically mounted detector on it shown on the right, ready to be lowered into the tank.}
\label{fig11}
\end{figure}

The dimensions of the tank allows multiple PMT modules to be tested underwater simultaneously, and also accommodate optical features studies from significant distances therein. 
A detector mounting system to hold the PMT-module or any other equipment is also prepared with aluminium extrusions, the
as shown in fig.\ref{fig11}. The mount is thoughtfully designed so as to support both the vertical and the horizontal mounting of the detector, or any target as heavy as 30~kg, without yielding to any bending.


\begin{figure}[b!t]
\includegraphics[width=1.\columnwidth]{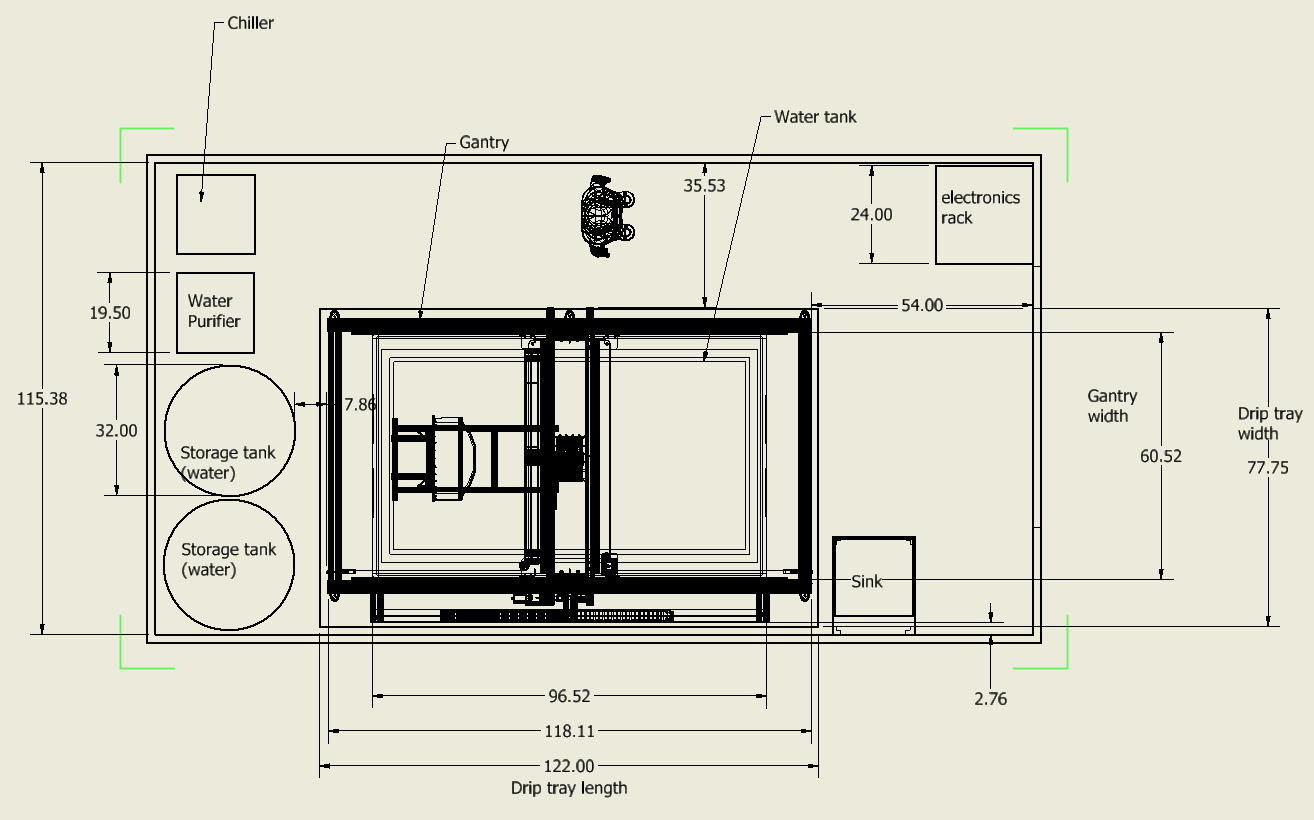}
\includegraphics[width=1.\columnwidth]{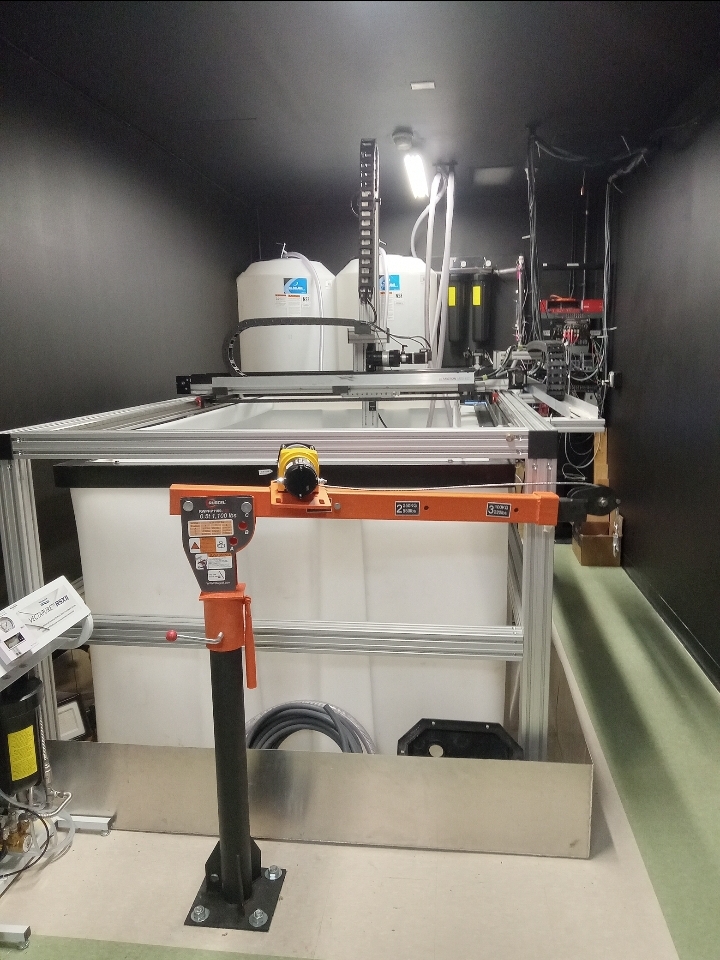}
    \caption{Top: Schematics of the top-view of the UWDTF facility. Bottom: A picture of the UWDTF facility at its first test run, after all prime components installed.}
\label{fig12}
\end{figure}

The top view schematics of the entire facility with almost all the necessary constituents is shown in fig.\ref{fig12}-top, while the bottom picture in fig.\ref{fig12} shows a photograph of the facility, all completed and at its first test run of all the components in 2023.

A crane has also been installed in the facility to aid in the lifting, lowering and moving of the objects and equipments in and out of the tank. 
The walls and the ceiling of the facilty are all painted black to ensure minimal light reflections during the dark room tests. The tank is layered with black tyvek sheets as and when necessary to serve the requirements of the experimental environment. The tests with the PMTs need to be done with absolute darkness in the room, while for the photogrammetric calibration studies, a diffused light source is preferred for some cases. 

\begin{figure}[b!t]
 \centering   
   \includegraphics[width=1.\columnwidth]{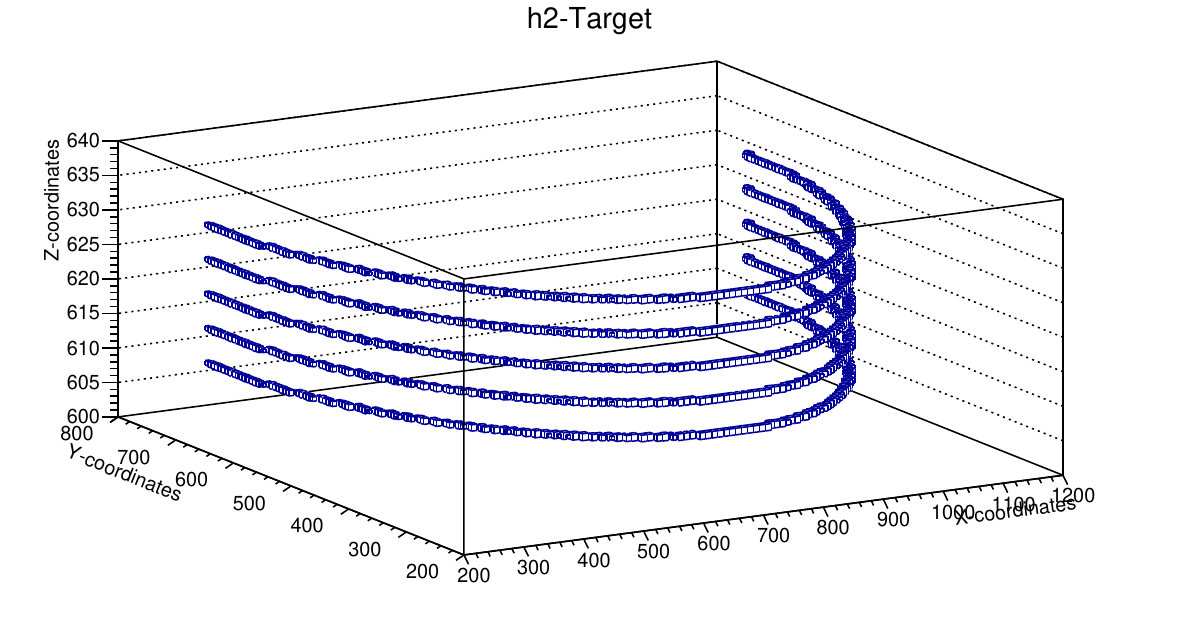}
    \caption{Cylindrically scanned coordinates by the Gantry arm around the centre of the arc, which in this case is the middle of one of the tank bottom edges.}
\label{fig13}
\end{figure}

The gantry ensures a smooth scanning of either the source or the detector attached to it as the payload around the detector or the source, respectively, inside the tank. One of the example cases is shown in fig.\ref{fig13}, where each point shows the position of the payload around the centre of the arc.

\section{Underwater Camera-Calibration Tests}

Several research and development activities are planned, being done and more to be realised at this multi-purpose Under-Water Dark-room Test Facility (UWDTF) built at the University of the Winnipeg. The primary motivation behind building this facility was to test the calibration apparatus to be used in the water Cherenkov detectors at the WCTE, IWCD, and the HyperK-FD. Cameras are a part of the calibration system to be used therein, and need an initial optical calibration to extract the camera intrinsic parameters in the underwater environment. The cameras will be installed in the water Cherenkov detectors and will aid the photogrammetric calibration methods of the detector.


The process of Camera Calibration involves determining the intrinsic parameters of a camera, like the focal length and principal point of the lens system, and is a standard procedure followed by many, like \cite{calib1, calib2, kannala, puigi,  scara1, steffen}. However, determination of 
 instrinsic parameters of a camera system inside water is one of its kind and has never been presented before. An optical medium of refractive index higher than that of air, that is, water, followed by an acryllic layer and a thin layer of air in front of the lens, in combination with the existing set of lenses inside the camera, prepare an altered effective lensing system, a significantly modified configuration than that of the camera placed in the air alone. The deployment of the gantry and the pan-tilt system further aids in reducing the errors from the most common sources in such camera calibration procedures, like relative displacement of the target with respect to the camera, due to unstability issues, and ensures several non-repetitive images to be taken according to a pre-mapped coordinates scan.   Therefore, the camera-calibration underwater referred to here, actually presents the calibration of a novel and unique lensing system in effect, and to be used by the internationally collaborated particle physics experiments in their huge water cherenkov detectors.

\begin{figure}[b!t]
 \centering   
   \includegraphics[width=1.\columnwidth]{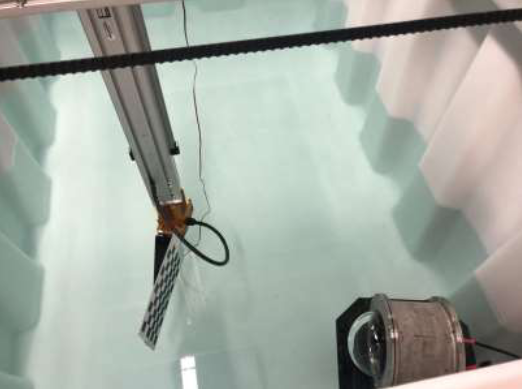}
    \caption{Camera and checkerboard target placed inside the tank with water. The target is attached to the axle of the pan-tilt system attached to the end of the gantry-arm.}
\label{fig14}
\end{figure}

The camera is packed in a waterproof steel housing with an acrylic dome towards the lens-side of the camera, and wired feedthrus at the posterior side of the chamber, to aid in recording images of the target with the camera, while both  underwater. A checkerboard pattern is printed on a waterproof metal board and attached to the end of the Gantry-arm, which will move around the camera positioned at the bottom of the tank, as shown in fig.\ref{fig14}.

\begin{figure}[b!t]
 \centering   
   \includegraphics[width=1.\columnwidth]{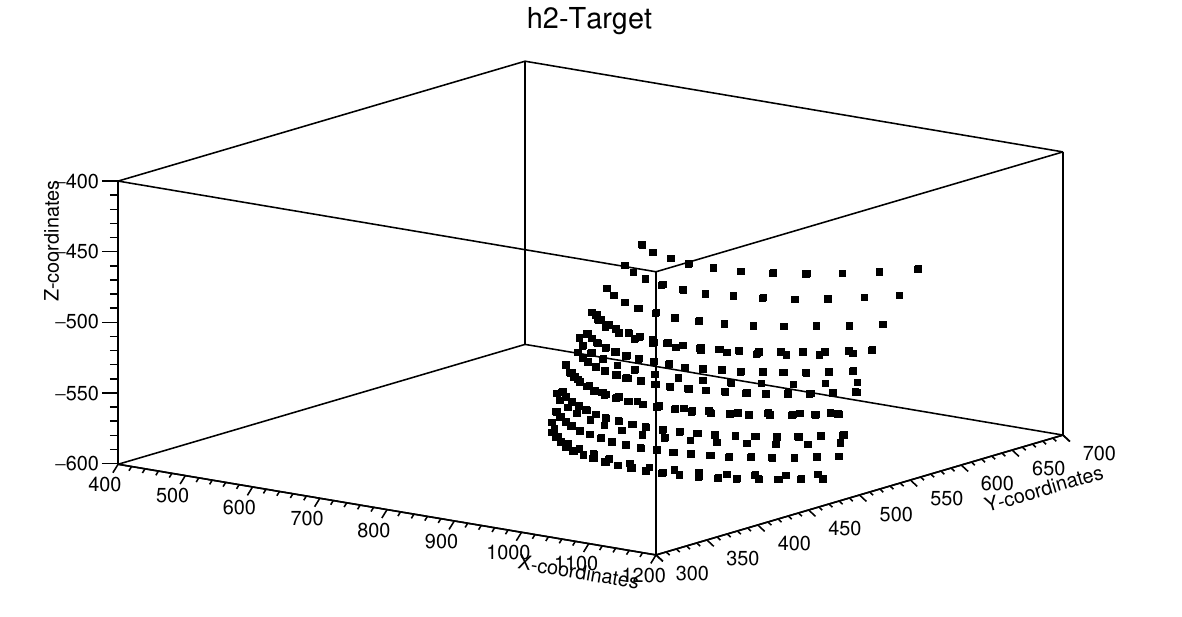}
    \caption{Spherically scanned coordinates by the Gantry arm around the centre of the arc, the centre in this case being the near-corner of one of the tank bottom edges. }
\label{fig15}
\end{figure}

Images have been recorded with various configurations with and without water, and scans done in cylindrical or spherical arc patterns for different tests and measurements. The different scenarios are being analysed and will be presented in future. The case of measurement discussed here is an exemplenary one and used the spherical scan under water as shown in the fig.\ref{fig15}.

\begin{figure}[b!t]
 \centering
   \includegraphics[width=1.\columnwidth]{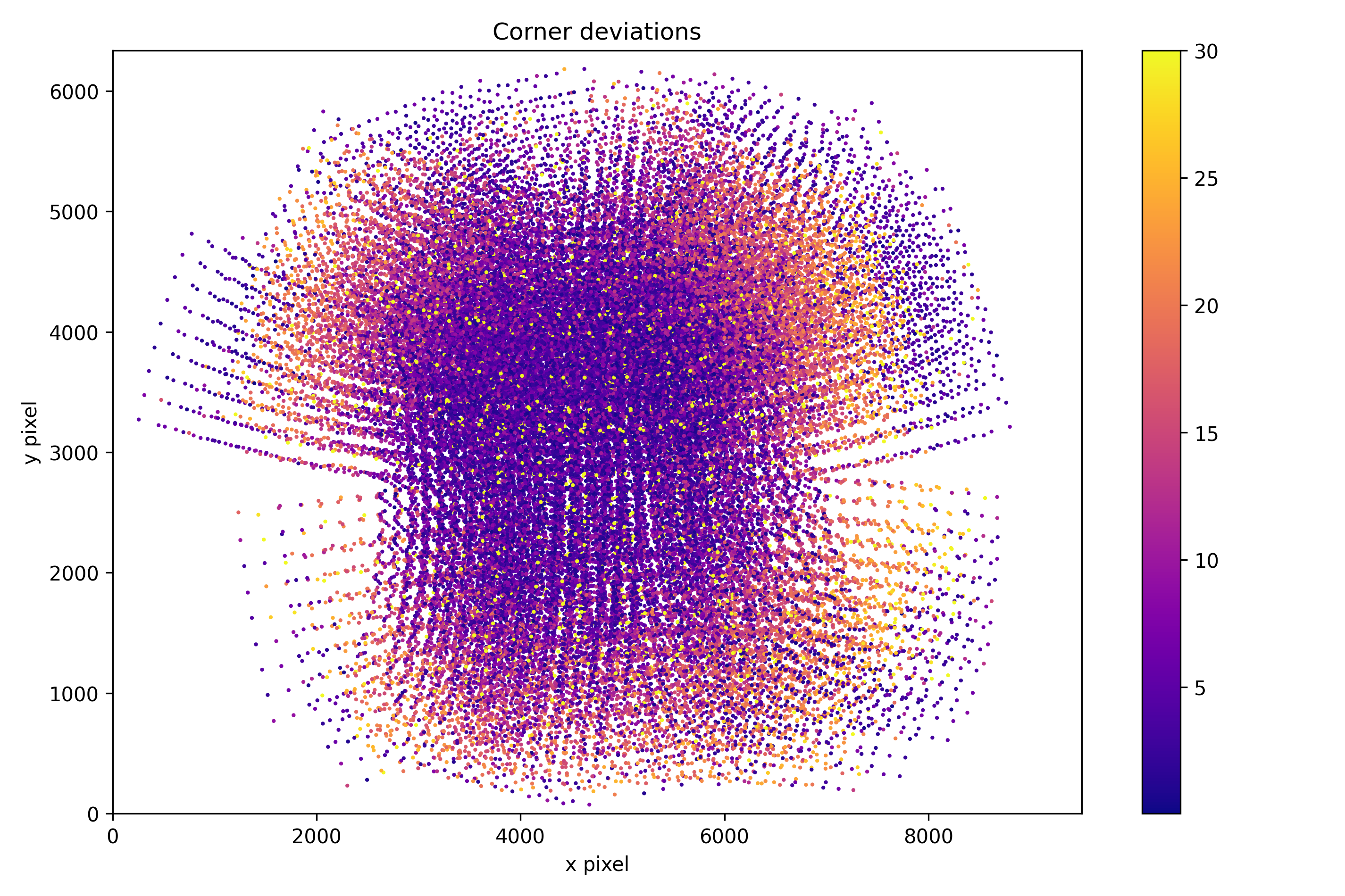}
    \caption{Corner deviations measured at each point of the calibration target.   }
\label{fig17}
\end{figure}

\begin{figure}[b!t]
 \centering
   \includegraphics[width=1.\columnwidth]{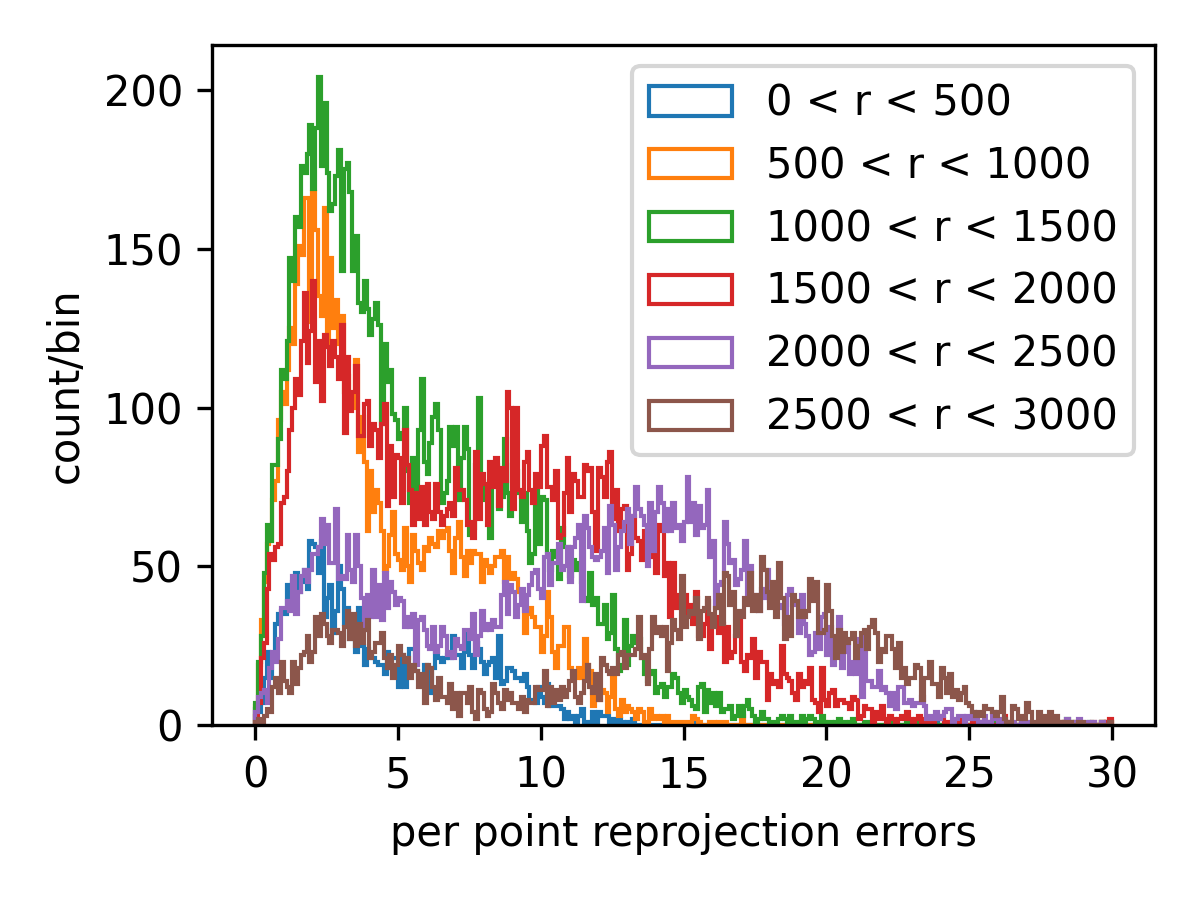}
    \caption{Corner deviation values at each point with respect to its radial distance from the principal axis.}
\label{fig18}
\end{figure}

The scanned coordinates as shown, cover the maximum  phase space or field of view of the camera, and the camera was rotated by 90$^\circ$ by tipping it on its side inside the tank, after each set of scan. The process was repeated four times and all sets of the images at each target position were recorded for analyses. The level of the water in the tank and the lighting inside the facility were adjusted, so that the images of the checkerboard pattern were clear enough and devoid of reflections. The images were then analysed using the open-CV library \cite{ocv}, using the fish-eye-lens model \cite{kannala}. The corners of the checkerboard pattern are located, followed by determining the  camera intrinsic and extrinsic parameters from the several views of the pattern. These parameters are then used to reproject the locations of the corners back onto the image plane. The process is repeated so as to minimize the errors in the reprojection. The parameters are hence obtained, and the deviations in each of the projected corner locations from that in the original image plane, as determined are shown in the fig.~\ref{fig17}, and
 are found to be mostly less than 5 pixels at the central and extreme regions of the image phase space, while peaking at around 2.5 pixels. However, many points in the intermediate region tend to have a higher deviation in the range $\sim$10-25 pixels, as can be seen in the plot. 
 The orientation of these values with respect to the distance from the centre of the image plane, i.e. the radial distance from the principal axis is shown in fig.~\ref{fig18}, for a better understanding.


The focal lengths along X and Y-directions are found to be 3.292$\times10^3$ and 3.290$\times10^3$ pixel units respectively, while the principal point is found to be at (x=4.662$\times10^3$ , y=3.092$\times10^3$) pixel units, 
which matches close to the mid-point of the array of the pixels (9504$\times$6336), affirming the spherical symmetry of the lens system.

\section{Conclusion}
An underwater dark-room research facility has been built in the University of Winnipeg, equipped with a 5-axes gantry system, functioning quite or mostly equivalent to the more expensive option of any robotic arm facility that may be found elsewhere. It rather includes much larger ranges of movement, not only in air but also underwater.  First of its kind, underwater calibration of cameras to determine the intrinsic parameters by a steady scan and vast number of data points, was successfully accomplished with minimal error introduction in the process, and the results from one of them presented in this paper. The detailed features in the results, as observed,  could be realised owing to the perfectly smooth  procedure of data-taking made possible by this newly built Facility and its elements, at the University of Winnipeg. Last but not the least, this is one of the many research projects that is/will be accommodated by this facility, providing the breeding place of more valuable research and studies.

\section*{Acknowledgement}
We're extremely grateful for the funding received from the CFI and the Research-Manitoba Funding Organisation, without which this project could not have been realisized.


\end{document}